\documentclass[prl,twocolumn,showpacs,superscriptaddress]{revtex4-1}

\usepackage[utf8]{inputenc}
\usepackage[english]{babel}
\usepackage[T1]{fontenc}

\usepackage{graphicx}
\usepackage{amsmath}
\usepackage{amsfonts}
\usepackage{amssymb}
\usepackage[breaklinks=true,colorlinks]{hyperref}

\usepackage{color}

\usepackage{lipsum}

\newcommand{\cf}{cf.\ }

\newcommand{\ie}{i.e.\ }

\newcommand{\coloneq}{\mathrel{\mathop:}=}
\newcommand{\dd}{\mathrm{d}}

\newcommand{\ket}[1]{\left|{#1}\right\rangle}
\newcommand{\ketbra}[2]{\left|{#1}\middle\rangle\middle\langle{#2}\right|}
\newcommand{\bkew}[3]{\left\langle{#1}\middle|{#2}\middle|{#3}\right\rangle}

\newcommand{\Tr}{\operatorname{Tr}}
\newcommand{\ew}[1]{\left\langle{#1}\right\rangle}

\newcommand{\kB}{k_\mathrm{B}}
\newcommand{\sminus}{\sigma_-}
\newcommand{\splus}{\sigma_+}

\newcommand{\indexc}{\mathrm{c}}
\newcommand{\indexh}{\mathrm{h}}
\newcommand{\Gc}{G_\indexc}
\newcommand{\Gh}{G_\indexh}
\newcommand{\betaeff}{\beta_\mathrm{eff}}
\newcommand{\betac}{\beta_\indexc}
\newcommand{\betah}{\beta_\indexh}
\newcommand{\Teff}{T_\mathrm{eff}}
\newcommand{\Tc}{T_\indexc}
\newcommand{\Th}{T_\indexh}
\newcommand{\Jc}{J_\indexc}
\newcommand{\Jh}{J_\indexh}
\newcommand{\psib}{\psi_\mathrm{b}}
\newcommand{\psid}{\psi_\mathrm{d}}

\begin{document}

\title{Power enhancement of heat engines via correlated thermalization in multilevel systems}

\author{David Gelbwaser-Klimovsky}
\thanks{These authors contributed equally to this work.}
\affiliation{Department of Chemical Physics, Weizmann Institute of Science, Rehovot~7610001, Israel}

\author{Wolfgang Niedenzu}
\thanks{These authors contributed equally to this work.}
\affiliation{Department of Chemical Physics, Weizmann Institute of Science, Rehovot~7610001, Israel}

\author{Paul Brumer}
\affiliation{Chemical Physics Theory Group, Department of Chemistry and Centre for Quantum Information and Quantum Control, University of Toronto, Ontario~M5S~3H6, Canada}

\author{Gershon Kurizki}
\email{gershon.kurizki@weizmann.ac.il}
\affiliation{Department of Chemical Physics, Weizmann Institute of Science, Rehovot~7610001, Israel}

\begin{abstract}
  We analyze a heat machine based on a periodically-driven quantum system permanently coupled to hot and cold baths. It is shown that the maximal power output of a degenerate $V$-type three-level heat engine is that generated by two independent two-level systems. For $N$ levels, this maximal enhancement is $(N-1)$-fold. Hence, level degeneracy is a thermodynamic resource that may effectively boost the power output. The efficiency, however, is not affected. We find that coherence is not an essential asset in multilevel-based heat machines. The existence of multiple thermalization pathways sharing a common ground state suffices for power enhancement.
\end{abstract}

\date{\today}

\maketitle

\paragraph{Introduction}

The rapport between quantum mechanics and thermodynamics is still an open problem~\cite{gemmerbook,kosloff2013quantum}. Its technological and fundamental implications have motivated numerous proposals of heat machines based on quantum systems~\cite{geva1996quantum,velasco1997new,palao2001quantum,quan2007quantum,birjukov2008quantum,allahverdyan2010optimal,esposito2010quantum,jahnke2010quantum,linden2010how,brunner2012virtual,esposito2012stochastically,gieseler2012subkelvin,kolar2012quantum,levy2012quantum,levy2012quantumrefrigerators,quan2012validity,correa2013performance,delcampo2013more,gelbwaser2013minimal,venturelli2013minimal}. Two main issues underlie such proposals: What are the bounds on the performance of quantum heat machines, \ie their power output and efficiency~\cite{gemmerbook,scovil1959three,geusic1959three,geusic1967quantum,kosloff2013quantum,alicki2014quantum}, and what thermodynamic properties (or resources) of quantum systems determine these bounds~\cite{boukobza2013breaking,gelbwaser2013work,gelbwaser2014heat,rossnagel2014nanoscale}? A pioneering approach addressing these issues~\cite{scully2002extracting,scully2011quantum} has identified steady-state coherence~\cite{agarwal2001quantum,kozlov2006inducing} between the levels of a quantum system as a thermodynamic resource.

\par

\begin{figure}
 \centering
 \includegraphics[width=\columnwidth]{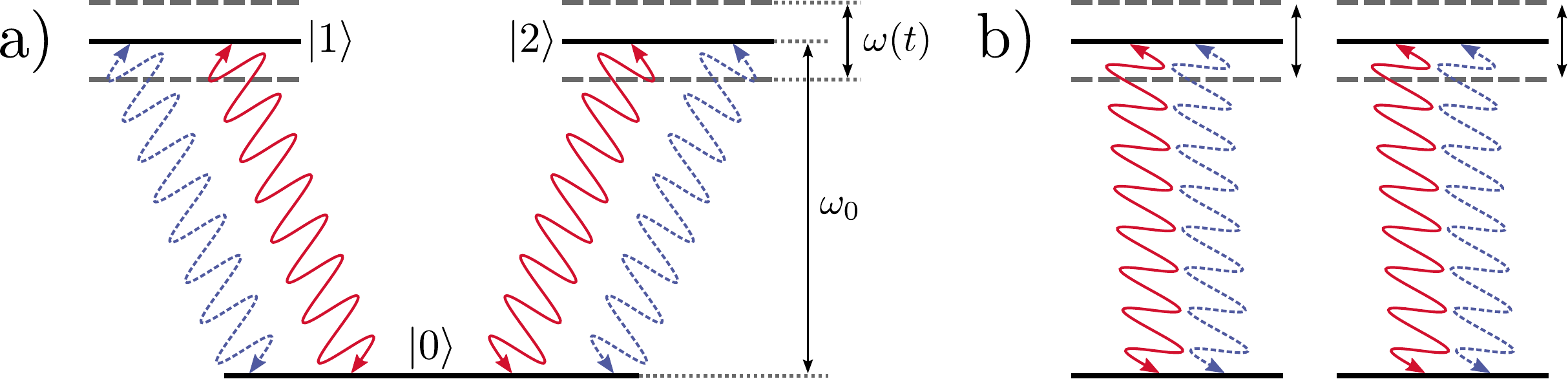}
 \caption{(Color online) a) Multilevel heat machine: A nearly-degenerate $V$-type three-level system whose upper levels undergo periodic modulation while simultaneously interacting with a cold and a hot bath. The ground and excited states are incoherently populated by absorption of quanta from and (spontaneous and stimulated) emission to these baths (dotted blue: cold bath, solid red: hot bath). b) Comparison to two independent two-level systems subject to the same environment and modulation.}\label{fig_system}
\end{figure}

Here we wish to elucidate these issues from first principles. We resort to a general framework for the description of a steady-state continuous heat machine based on a periodically-driven quantum system permanently coupled to hot and cold baths~\cite{kolar2012quantum,gelbwaser2013minimal}. This framework enforces consistency with the first and second laws by constructing appropriate heat currents flowing between the baths via the system~\cite{alicki2012periodically,gelbwaser2013minimal}. To account for the possible r\^ole of coherences we extend this framework, hitherto applied to a two-level system (TLS)~\cite{kolar2012quantum,gelbwaser2013minimal}, to an analogous heat engine based on a multilevel system, in particular a $V$-type \emph{three-level system} as depicted in Fig.~\ref{fig_system}a. The performance is compared to two independent TLS heat engines (\cf Fig.~\ref{fig_system}b), where steady-state coherence is absent. We show that steady-state coherence does not affect the efficiency, nor does maximal power boost necessarily require coherence.

\paragraph{Model and steady-state treatment}

We consider a $V$-type three-level system with degenerate excited states $\ket{1}$ and $\ket{2}$, ground state $\ket{0}$ and transition frequency $\omega_0$. To operate a heat machine, we simultaneously connect this system to two (hot and cold) baths and a ``piston'', here realized by a periodic modulation of the upper levels~\cite{kolar2012quantum,gelbwaser2013minimal} that results in a periodic transition frequency $\omega_0+\omega(t),$ with $\omega\left(t+\frac{2\pi}{\Omega}\right)=\omega(t)$, where $\Omega$ denotes the modulation rate. The system--bath interaction is described by the following generic Hamiltonian in the rotating wave approximation,
\begin{equation}\label{eq_H_SB}
  H_\mathrm{SB}=\sum_{j=1}^2\sum_{i\in\{\indexc,\indexh\}}\left(\splus^j\otimes \mathbf{d}_{j}\cdot\mathbf{B}_i + \sminus^j\otimes \mathbf{d}_{j}^*\cdot\mathbf{B}_i^\dagger\right),
\end{equation}
where $\splus^j\coloneq\ketbra{j}{0}$, $\sminus^j\coloneq\ketbra{0}{j}$, $\mathbf{d}_{j}$ is the transition dipole between the excited state $\ket{j}$ and the ground state $\ket{0}$, and $\mathbf{B}_i$ is the hot ($\indexh$) or cold ($\indexc$) bath operator. For simplicity we here restrict the treatment to real dipoles of equal strength, $d=|\mathbf{d}_{1}|=|\mathbf{d}_{2}|$. These transition dipoles need not to be parallel (aligned), as discussed below. 

\par

As detailed in~\cite{alicki2012periodically} the periodicity of the modulation implies that the dynamics of the system's density matrix in the interaction picture is governed by a linear combination of ``sub-bath'' Lindblad operators, i.e., operators associated with the two baths $i\in\{\indexc,\indexh\}$, evaluated at the harmonic sidebands $q=0,\pm1,\pm2,\dots$ of the modulation frequency $\Omega$. The master equation then reads
\begin{subequations}\label{eq_master_L}
  \begin{equation}\label{eq_master}
    \dot\rho=\sum_{q\in\mathbb{Z}}\sum_{i=\{\indexc,\indexh\}}\mathcal{L}_i^q\rho
  \end{equation}
  with the Liouvillian superoperators of the $(i,q)$ ``sub-baths''
  \begin{multline}\label{eq_L}
    \mathcal{L}_i^q\rho=P(q)\frac{G_i(\omega_0+q\Omega)}{2}\\
    \times\sum_{j=1}^2\left[\mathcal{D}(\sminus^j,\splus^j)+\sum_{j^\prime\neq j}\mathfrak{p}\mathcal{D}(\sminus^j,\splus^{j'})\right]\\
    +P(q)\frac{G_i(-\omega_0-q\Omega)}{2}\\
    \times\sum_{j=1}^2\left[\mathcal{D}(\splus^j,\sminus^j)+\sum_{j^\prime\neq j}\mathfrak{p}\mathcal{D}(\splus^j,\sminus^{j'})\right].
  \end{multline}
\end{subequations}
Here $P(q)$ is the weight of the $q$th harmonic (determined by the modulation form)~\cite{alicki2012periodically} and the dissipator reads $\mathcal{D}(a,b)\coloneq 2a\rho b-ba\rho-\rho ba$ for any operators $a,b$. The factors $G_i(\pm\omega)$ are the coupling spectra to the $i$th bath. They fulfill the KMS condition~\cite{breuerbook} $G_i(-\omega)=e^{-\beta_i\hbar\omega}G_i(\omega)$ 
with $G_i(\omega)=\gamma_i(\omega)\left(\bar{n}_i(\omega)+1\right)$, where $\gamma_i(\omega)$ is the frequency-dependent transition rate induced by the $i$th bath and $\bar{n}_i(\omega)$ denotes the corresponding number of thermal quanta at inverse temperature $\beta_i=1/\kB T_i$. The dissipators $\mathcal{D}(\sminus^j,\splus^j)$ and $\mathcal{D}(\splus^j,\sminus^j)$ describe emission and absorption (and hence population transfer) involving separate transitions ($\ket{1}\leftrightarrow \ket{0}$ and $\ket{2}\leftrightarrow \ket{0}$) via their common ground state. By contrast, $\mathcal{D}(\sminus^j,\splus^{j'\neq j})$ and $\mathcal{D}(\splus^j,\sminus^{j'\neq j})$ describe \emph{cross-correlations} between the two transitions, allowing for bath-induced quanta exchange between the two excited states and thereby generating coherences between these states. An important parameter is the dipole alignment factor $\mathfrak{p}\coloneq\frac{\mathbf{d}_{1}\cdot\mathbf{d}_{2}}{|\mathbf{d}_{1}||\mathbf{d}_{2}|}\equiv \cos\measuredangle(\mathbf{d}_{1},\mathbf{d}_{2})$. In the SI, the cases of parallel and anti-parallel dipoles are shown to be analogous, so that we can without loss of generality restrict the domain to $\mathfrak{p}\in[0,1]$.

\paragraph{Analysis}

The energy that is continuously exchanged between the three-level system and the heat baths is related, according to the \emph{first law}, to the power (the rate of work $W$ extracted by the piston) by~\cite{callenbook} $\dot{W}=-(\Jc+\Jh)$. This expression involves the sum of heat currents from both baths, which can be derived from the dynamical version of the \emph{second law}~\cite{kosloff2013quantum}. Their explicit expression for the $i$th bath ($i\in\{\indexc,\indexh\}$) is $J_i=\sum_{q}J_i^q$, where the heat current $J_i^q$ for the $q$th harmonic ``sub-bath'' ($q\in\mathbb{Z}$) in Eq.~\eqref{eq_L} reads~\cite{alicki2012periodically,kosloff2013quantum}
\begin{equation}\label{eq_J}
  J_i^q\coloneq-\frac{1}{\beta_i}\Tr\left[(\mathcal{L}_i^q\rho)\ln(\rho_i^q)\right].
\end{equation}
Here, $\rho^q_i$ denotes the \emph{local} steady state for a \emph{single} heat bath at temperature $T_i$ evaluated at the sideband $\omega_0+q\Omega$. We stress that the \emph{global} steady state $\rho$ ensures the correct description of heat transport in this correlated three-state system, avoiding inconsistencies with the second law due to the improper use of local variables, as discussed in~\cite{levy2014local}. Since every Liouvillian $\mathcal{L}_i^q$ in the master equation~\eqref{eq_master} has the same functional dependence~\eqref{eq_L} on the atomic operators, the correct global solution can be directly obtained from the local one.

\par

We here search for the steady-state solution of the master equation~\eqref{eq_master} and the resulting expressions for $J_{\indexh(\indexc)}$. At this point we are unaware of the bound for these currents and its dependence on alignment. These heat currents are therefore compared to the corresponding expressions $J_i^\mathrm{TLS}$ for a two-level system (TLS) with the same transition-dipole strength $d$ and modulated transition frequency $\omega_0+\omega(t)$~\cite{gelbwaser2013minimal}.

\par

The master equation~\eqref{eq_master} can be reduced to an analytically solvable inhomogeneous system of linear differential equations $\dot{\mathbf{x}}=\mathcal{A}\mathbf{x}+\mathbf{b}$ for the vector of matrix elements $\mathbf{x}\coloneq(\rho_{21},\rho_{12},\rho_{00},\rho_{22})^T$. This system of ordinary differential equations (ODEs) (see SI) describes two very distinct dynamical regimes corresponding to aligned and misaligned transition dipoles, as detailed in what follows.

\par

\begin{figure}
  \centering
  \includegraphics[width=\columnwidth]{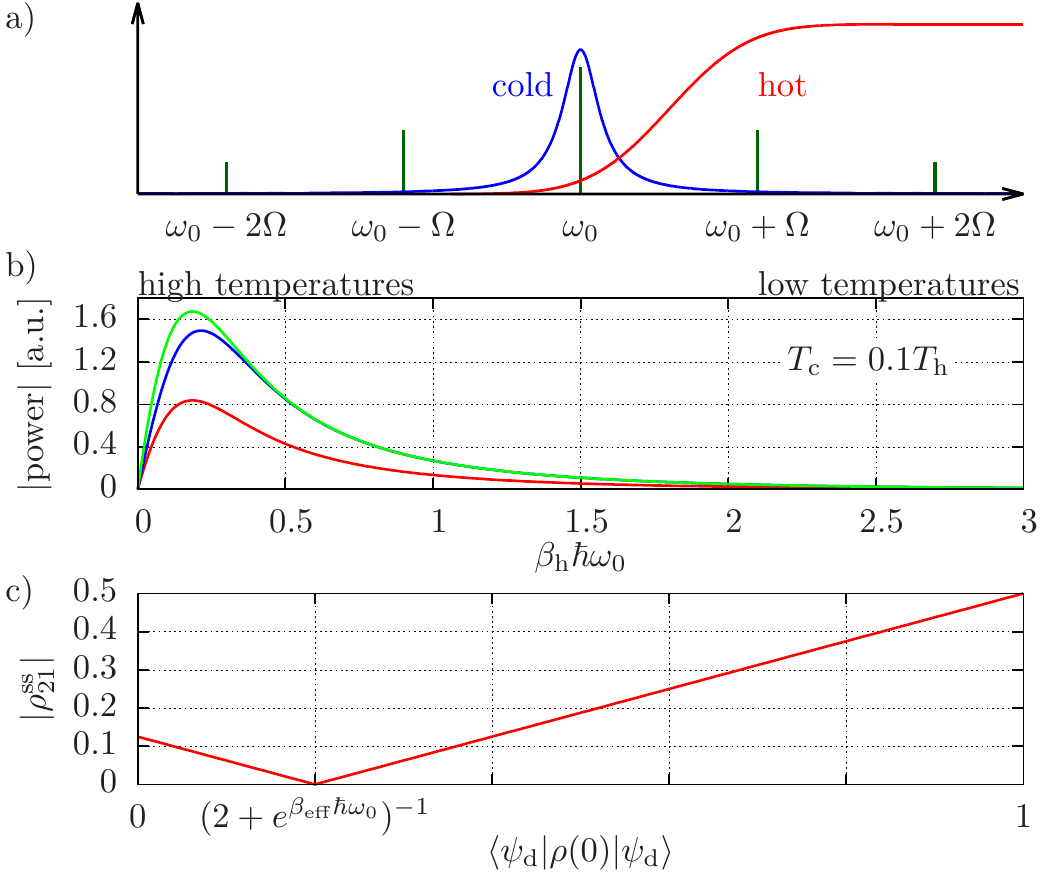}
  \caption{(Color online) a) ``Engineering'' of the effective temperature $\Teff$ by controlling the weights of harmonic sidebands (via the modulation) in the two bath spectra. b) Absolute value of the maximum power extraction (from bottom to top: TLS, non-aligned three-level system, aligned three-level system under optimal initial conditions) for $\Tc=0.1\Th$. c) Modulus $|\rho_{21}^\mathrm{ss}|$ of the steady-state coherence for parallel dipoles. Maximal power boost [occurring for small initial dark-state populations according to Eq.~\eqref{eq_power_aligned}] corresponds to relatively small steady-state coherences. The highest steady-state coherence is realized for a dark initial state, which yields zero power.}\label{fig_spectra_power_coherences}
\end{figure}

(i) Let us first consider the very general steady-state regime obtained for near-degenerate upper levels with similar transition frequencies $\omega_j$ ($j=1,2$) and \emph{misaligned transition dipoles}, $\mathfrak{p}\in[0,1)$. The three-level system then thermalizes to the \emph{diagonal} steady state (without coherences)
\begin{align}\label{eq_ode_solution_pneq1}
  \rho_{jj}^\mathrm{ss}&=\rho_{00}^\mathrm{ss}\frac{\sum_q\sum_i P(q) G_i(-\omega_j-q\Omega)}{\sum_q\sum_i P(q) G_i(\omega_j+q\Omega)}.
\end{align}
In what follows we neglect for simplicity the difference between the transition frequencies, \ie we set $\omega_j=\omega_0$. Then the cumbersome r.h.s.\ of Eq.~\eqref{eq_ode_solution_pneq1} is reduced to a Boltzmann factor for an effective inverse temperature $\betaeff$ defined by
\begin{equation}\label{eq_betaeff}
  e^{-\betaeff\hbar\omega_0}\coloneq\frac{\sum_q\sum_i P(q)G_i(-\omega_0-q\Omega)}{\sum_q\sum_i P(q)G_i(\omega_0+q\Omega)},
\end{equation}
which is bounded by the real inverse temperatures $\betah\leq\betaeff\leq\betac$ of the two baths. This effective temperature determines the steady-state populations of the periodically modulated system coupled to both baths. We can control $\betaeff$ by engineering the modulation Floquet coefficients $P(q)$ that determine the overlap of the sideband peaks ($q=\pm1,\pm2,\dots$) at the frequency harmonics $\omega_0+q\Omega$ with the response spectra $G_i(\omega)$ of the two baths, as sketched in Fig.~\ref{fig_spectra_power_coherences}a.

\par

Upon computing the heat currents~\eqref{eq_J}, we find that $\Jh$, $\Jc$ and the power $\dot W$ are modified (relative to their TLS counterparts~\cite{gelbwaser2013minimal}) by the same factor
\begin{equation}\label{eq_power_nonaligned}
  \frac{\dot{W}}{\dot{W}^\mathrm{TLS}}\equiv\frac{J_i}{J_i^\mathrm{TLS}}=2\frac{\rho_{00}^\mathrm{ss}}{\rho_{00}^\mathrm{TLS}}=2\frac{1+e^{-\betaeff\hbar\omega_0}}{1+2e^{-\betaeff\hbar\omega_0}}.
 \end{equation} 
This means that the power enhancement relative to a TLS heat machine is determined by the ratio of the steady-state ground-state population $\rho_{00}^\mathrm{ss}$ in the $V$-system to its TLS counterpart. Namely, in this fully thermalized incoherent regime the enhancement factor~\eqref{eq_power_nonaligned} only depends on the effective temperature~\eqref{eq_betaeff}.

\par

(ii) For fully degenerate excited states and \emph{aligned} dipole moments ($\mathfrak{p}=1$) we find  that $\det(\mathcal{A})=0$ in the ODE above, due to a singularity of the coefficient matrix $\mathcal{A}$ (see SI). This implies that an infinite number of steady-state solutions may exist. Indeed, in this regime the dynamics is constrained by the existence of a dark state which renders the steady-state solution dependent on the initial conditions (in agreement with the expression found for a single bath in~\cite{agarwal2001quantum}). The steady-state solution now depends on the overlap of the initial state $\rho(0)$ with the non-dark states (\ie the ground state and the bright state) of the full Liouvillian $\mathcal{L}$ in Eqs.~\eqref{eq_master_L}. The r\^ole of these states becomes apparent upon diagonalizing the steady-state solution that yields the populations (see SI)
\begin{subequations}\label{eq_eigenvalues}
  \begin{align}
    \rho_{\mathrm{b}\mathrm{b}}^\mathrm{ss}&=\frac{1}{1+e^{\betaeff\hbar\omega_0}}\left[\rho_{\mathrm{bb}}(0)+\rho_{00}(0)\right]\equiv e^{-\betaeff\hbar\omega_0}\rho_{00}^\mathrm{ss}\\
    \rho_{\mathrm{d}\mathrm{d}}^\mathrm{ss}&=\bkew{\psid}{\rho(0)}{\psid}
  \end{align}
\end{subequations}
in the basis spanned by $\{\ket\psib,\ket\psid,\ket0\}$. Here $\ket{\psib}\coloneq\frac{1}{\sqrt{2}}\left(\ket{1}+\ket{2}\right)$ and $\ket{\psid}\coloneq\frac{1}{\sqrt{2}}\left(\ket{1}-\ket{2}\right)$ denote the bright and dark states, respectively. Whilst the dark-state population cannot change, \ie it is a constant of motion (consistent with the one obtained in~\cite{kozlov2006inducing} for a single zero-temperature bath and external driving), the bright- and ground-state populations, $\rho_\mathrm{bb}$ and $\rho_{00}$, respectively, thermalize. The same results hold for anti-parallel dipoles upon interchanging the dark and the bright states.

\par

Proceeding as before in the non-aligned case, we find the power ratio
\begin{equation}\label{eq_power_aligned}
  \frac{\dot{W}}{\dot{W}^\mathrm{TLS}}\equiv\frac{J_i}{J_i^\mathrm{TLS}}=2\frac{\rho_{00}^\mathrm{ss}}{\rho_{00}^\mathrm{TLS}}=2\left[\rho_{\mathrm{bb}}(0)+\rho_{00}(0)\right].
\end{equation}
Hence, the power (as well as the heat currents) in the aligned regime are enhanced relative to their TLS counterparts by at most a factor of two, just as in the misaligned regime [Eq.~\eqref{eq_power_nonaligned}]. Yet, contrary to the latter, the ratio~\eqref{eq_power_aligned} does \emph{not} depend on the bath spectra but solely on the initial populations of the non-dark states. Enhancement in~\eqref{eq_power_aligned} requires $\rho_{\mathrm{bb}}(0)+\rho_{00}(0)>\frac{1}{2}$, or, equivalently, $\bkew{\psid}{\rho(0)}{\psid}<\frac{1}{2}$, \ie at least half of the initial-state population has to be non-dark. Maximal enhancement occurs when the \emph{initial} state is \emph{amenable to full thermalization}, \ie it is non-dark.
\par
For a given \emph{initial} ground-state population $\rho_{00}(0)$, the states orthogonal to the dark subspace are characterized by $\rho_{11}(0)=\rho_{22}(0)=\rho_{21}(0)=\frac{1}{2}\left[1-\rho_{00}(0)\right]$. These are the states with the maximally allowed \emph{modulus} of the $\rho_{21}(0)$ coherence (for a fixed ground-state population) \emph{and} the correct phase. We have plotted the maximum power output under sinusoidal modulation for a TLS, a non-aligned, and an aligned $V$-system in Fig.~\ref{fig_spectra_power_coherences}b. The spectra are chosen as in~\cite{gelbwaser2013minimal} such that only $\Gc(\omega_0)$ and $\Gh(\omega_0+\Omega)$ contribute (as sketched in Fig.~\ref{fig_spectra_power_coherences}a) and the modulation frequency has been tuned to the value maximizing the power output.
\par
We stress that a non-dark initial state does not correspond to a steady state with maximal coherence $|\rho_{21}^\mathrm{ss}|$. In fact, the coherence $|\rho_{21}^\mathrm{ss}|$ is maximized for an initial dark state, which does not exchange energy with the baths and gives zero power, see Fig.~\ref{fig_spectra_power_coherences}c.

\par

It is natural to ask: How much initial overlap with the dark state is allowed such that the aligned configuration still outperforms its misaligned counterpart? The answer is, for $\bkew{\psid}{\rho(0)}{\psid}<\frac{1}{2+e^{\betaeff\hbar\omega_0}}$. The latter becomes an equality for the overlap for which the coherences vanish even in the aligned case (see Fig.~\ref{fig_spectra_power_coherences}c).

\par

Let us now compare the enhancement factors~\eqref{eq_power_nonaligned} and \eqref{eq_power_aligned} for the misaligned ($\mathfrak{p}<1$) and aligned ($\mathfrak{p}=1$) regimes. Their ratio is determined by their respective steady-state populations in the ground state, which is directly related to the power or heat-current ratio via
\begin{equation}\label{eq_heat_current_ratio}
  \frac{{\dot W}^{\mathfrak{p}=1}}{{\dot W}^{\mathfrak{p}<1}}\equiv\frac{J_i^{\mathfrak{p}=1}}{J_i^{\mathfrak{p}<1}}=\frac{\rho_{00}^{\mathfrak{p}=1}}{\rho_{00}^{\mathfrak{p}<1}}.
\end{equation}
We consider this ratio in two limiting cases:

\par

(i) As $\betaeff\rightarrow 0$ (high effective temperature) the thermalized state corresponds to equipartition amongst the levels. For parallel dipoles, the thermalized three-level system behaves as a TLS (formed by the ground and the bright states) with an effective dipolar transition enhanced by the number of thermalization pathways, in this case two. Hence, in steady state half of the population is found in level $\ket{0}$ (if the initial state had no dark component). For misaligned dipoles, by contrast, thermal equilibrium corresponds to the equipartition amongst the three levels $\ket{1}$, $\ket{2}$ and $\ket{0}$. Consequently, only a third of the population is found in the ground state. The $3/2$ ratio of the respective ground-state populations according to Eq.~\eqref{eq_heat_current_ratio} explains the ratio of the maximal enhancement factors in the aligned and misaligned regimes at high $\Teff$. 

\par

(ii) For large $\betaeff$, \ie low $\Teff$, however, Eq.~\eqref{eq_heat_current_ratio} implies that the maximal enhancement for misaligned dipoles coincides with its counterpart for aligned dipoles (the latter is maximized for an initial state perpendicular to the dark state), since only $\ket{0}$ is then appreciably populated in either regime. 

\par

\begin{figure}
 \centering
 \includegraphics[width=\columnwidth]{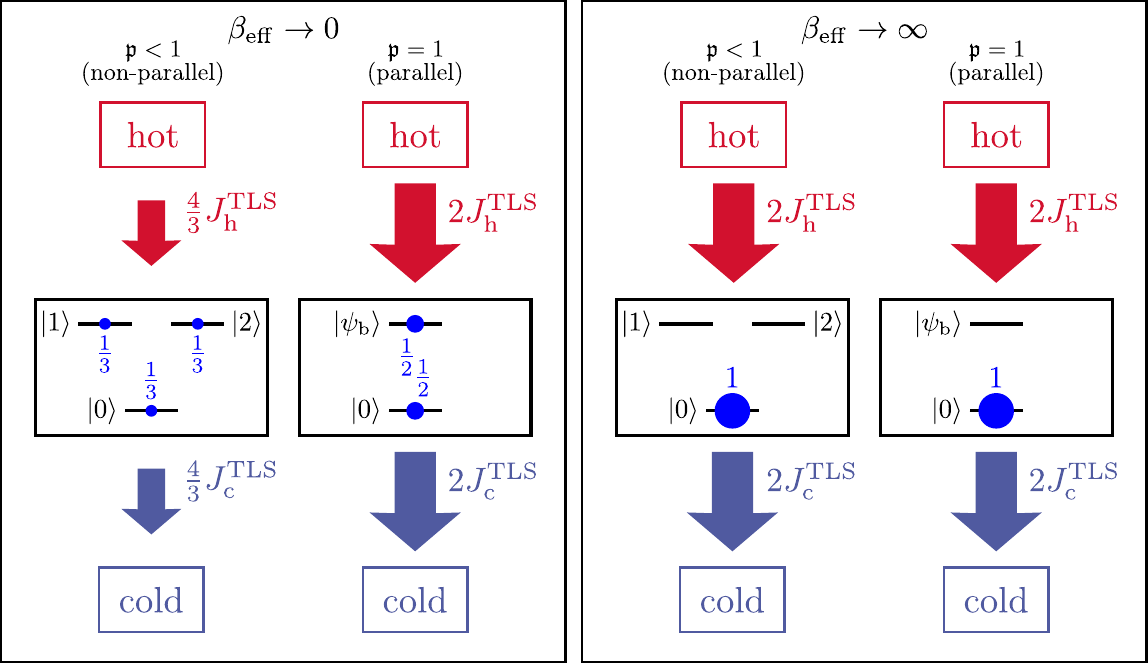}
 \caption{(Color online) Limiting regimes of the heat currents for high (left panel) and low (right panel) effective temperatures. For the parallel case we assume an optimal initial state (orthogonal to the dark state). The heat currents are related to the steady-state population of the ground state via Eq.~\eqref{eq_heat_current_ratio}.}\label{fig_heat_currents}
\end{figure}

Both regimes still retain the maximal enhancement factor of $2$, because of their double thermalization pathways instead of one for a genuine TLS. We have summarized these results in Fig.~\ref{fig_heat_currents}. A beneficial influence of alignment on power output is only expected for effective temperatures $\kB\Teff\gtrsim\hbar\omega_0/10$. For optical transitions this corresponds to a few hundred Kelvin, whereas for microwave transitions the benefit of alignment is already expected for a few hundred milli-Kelvin.

\par

\par

The foregoing results will be generalized to $N$-level systems in a forthcoming article. The maximal enhancement factor of the power output relative to a TLS is then found to be $N-1$, which is the number of thermalization pathways. This can be clearly seen in the case of $N-1$ fully aligned, degenerate upper states, yielding the generalization of~\eqref{eq_power_aligned} as follows,
\begin{equation}
  \frac{\dot{W}}{\dot{W}^\mathrm{TLS}}=(N-1)\left[1-\ew{\Pi_\mathrm{d}}_{\rho(0)}\right],
\end{equation}
where $\ew{\Pi_\mathrm{d}}_{\rho(0)}$ is the projection of the initial state $\rho(0)$ onto the dark subspace.

\paragraph{Realizations} 

$V$-systems with degenerate upper states are commonplace in atoms devoid of hyperfine interactions, e.g., Hg. The simultaneous coupling of such systems to hot and cold baths with controlled spectra can realize the misaligned case.

\par

The case of degenerate upper states and parallel transition dipoles (which, as discussed, is beneficial for power enhancement \emph{only for high} $\Teff$), is obtainable only for transitions between a lower state with angular momentum $l$ and magnetic number $m$ and degenerate upper states with the \emph{same} $m$. In atomic degenerate $V$-systems such parallel transition dipoles are forbidden by selection rules. Their are, however, proposals that dressed states may effectively realize such systems~\cite{ficek2002quantum}. Molecules may be another promising possibility due to their rich level structure involving vibrational degrees of freedom~\cite{[{See supplementary information in }] tscherbul2014long}.

\paragraph{Discussion}

Regardless of the transition-dipole alignment or small deviations from degeneracy, the maximally enhanced power output of a near-degenerate $V$-system heat engine is that generated by \emph{two independent} two-level systems. But as shown, there are conditions where degenerate levels sharing a common ground state can reach the same enhancement. For a generalization to $N$ levels, this maximal enhancement is ($N-1$)-fold. Hence, level degeneracy is found to be a \emph{thermodynamic resource} that may effectively boost the power output. Yet, it does not affect the efficiency: Since the same modifying factors [Eqs.~\eqref{eq_power_nonaligned} and \eqref{eq_power_aligned}] are obtained for the heat currents and the power, the \emph{efficiency } $\eta\coloneq-\dot{W}/J_\mathrm{\indexh}$ of the degenerate multilevel heat machine is the \emph{same as for a two-level system.} Hence, the same universal dependence of the efficiency on the modulation rate found in~\cite{gelbwaser2013minimal} holds for the present system. In particular, as the heat currents~\eqref{eq_J} (by construction) fulfill the second and the first laws, they adhere to the Carnot bound~\cite{alicki2014quantum} (this may not be true when the modulation is replaced by a quantized piston~\cite{gelbwaser2013work,gelbwaser2014heat}).

\par

At effective temperatures significantly larger than the level spacing, the aligned regime, where steady-state coherences may play a r\^ole, can outperform all misaligned cases. However, even when the aligned regime is advantageous, we may not attribute the power enhancement to steady-state coherence but rather to the ability of the initial state to completely thermalize. Steady-state coherence is the consequence of this thermalization but not its cause. We therefore conclude that coherence is not an essential asset in multilevel-based heat machines. The existence of multiple thermalization pathways sharing a common ground state suffices for power enhancement. 

\begin{acknowledgments}
  This work has been supported by the ISF, BSF and AERI (G.K.) and CONACYT (D.G.-K.).
\end{acknowledgments}

\clearpage

\begin{widetext}

\section{Supplemental Material}

\subsection{Linear ODE system}

As mentioned in the main text, the master equation can be mapped onto the linear system of ordinary differential equations
\begin{equation}\label{eq_odes}
 \frac{\dd}{\dd t}\mathbf{x} = \mathcal{A}\mathbf{x}+\mathbf{b},
\end{equation}
with
\begin{equation}
 \mathbf{x}\coloneq
 \begin{pmatrix}
 \rho_{21} \\
 \rho_{12} \\
 \rho_{00} \\
 \rho_{22} 
 \end{pmatrix}
 ,
\end{equation}
the coefficient matrix
\begin{subequations}
 \begin{equation}\label{eq_A}
 \mathcal{A}\coloneq \frac{1}{2}\sum_{q\in\mathbb{Z}}\sum_{i\in\{\indexc,\indexh\}}G_i(\omega_0+q\Omega)
 \begin{pmatrix}
  -2 & 0 & \mathfrak{p} \left(1+2 e^{-\betaeff\hbar\omega_0}\right) & 0 \\
  0 & -2 & \mathfrak{p} \left(1+2 e^{-\betaeff\hbar\omega_0}\right) & 0 \\
  2\mathfrak{p} & 2\mathfrak{p}  & -2\left(1+2e^{-\betaeff\hbar\omega_0}\right) & 0 \\
  -\mathfrak{p} & -\mathfrak{p}  & 2 e^{-\betaeff\hbar\omega_0} & -2
 \end{pmatrix}
 \end{equation}
 and the inhomogeneity
 \begin{equation}
 \mathbf{b}\coloneq \frac{1}{2}\sum_{q\in\mathbb{Z}}\sum_{i\in\{\indexc,\indexh\}}G_i(\omega_0+q\Omega)
 \begin{pmatrix}
  -\mathfrak{p} \\ -\mathfrak{p} \\ 2 \\ 0
 \end{pmatrix}
 .
 \end{equation}
\end{subequations}
Note that the coherences between the ground and the excited states ($\rho_{10}$ and $\rho_{20}$) do not appear as they follow a decoupled dynamics (leading to vanishing steady-state values).

\par

The determinant of the coefficient matrix~\eqref{eq_A} reads
\begin{equation}
  \det(\mathcal{A})=\left[\frac{1}{2}\sum_{q\in\mathbb{Z}}\sum_{i\in\{\indexc,\indexh\}}G_i(\omega_0+q\Omega)\right]^416\left(1+2e^{-\betaeff\hbar\omega_0}\right)\left(1-\mathfrak{p}^2\right)
\end{equation}
and clearly reveals a singularity for $\mathfrak{p}=\pm1$, \ie for parallel and anti-parallel dipoles. The latter two cases thus behave likewise, which justifies w.l.o.g.\ to restrict the domain of $\mathfrak{p}$ to $[0,1]$ as done in the main text. 

\section{Steady-state solution for parallel dipoles}

The steady-state solution of Eq.~\eqref{eq_odes} for $\mathfrak{p}=1$ reads
\begin{subequations}
 \begin{align}
   \rho_{00}^\mathrm{ss}&=\frac{1}{1+e^{-\betaeff\hbar\omega_0}}\left[1-\bkew{\psid}{\rho(0)}{\psid}\right]\\
   \rho_{22}^\mathrm{ss}&=\frac{1}{2}\frac{1}{1+e^{\betaeff\hbar\omega_0}}\left[1+\bkew{\psid}{\rho(0)}{\psid}e^{\betaeff\hbar\omega_0}\right]\\
   \rho_{11}^\mathrm{ss}&=\frac{1}{2}\frac{1}{1+e^{\betaeff\hbar\omega_0}}\left[1+\bkew{\psid}{\rho(0)}{\psid}e^{\betaeff\hbar\omega_0}\right]\\
   \rho_{21}^\mathrm{ss}&=\frac{1}{2}\frac{1}{1+e^{\betaeff\hbar\omega_0}}\left[1-\bkew{\psid}{\rho(0)}{\psid}\left(2+e^{\betaeff\hbar\omega_0}\right)\right].
 \end{align}
\end{subequations}
Upon diagonalization one recovers the diagonal steady-state solution provided in the main text [Eqs.~\eqref{eq_eigenvalues}].

\section{Power for a TLS-based heat machine}

For completeness we provide here the explicit formula for a minimal heat machine based on a periodically-modulated TLS coupled to two baths (Ref.~\cite{gelbwaser2013minimal} in the main text),

\begin{equation}
  \dot{W}^\mathrm{TLS}=-\sum_{i\in\{\indexc,\indexh\}}J_i^\mathrm{TLS}=\sum_{q\in\mathbb{Z}}\sum_{i\in\{\indexc,\indexh\}}\frac{\hbar\left(\omega_0+q\Omega\right)P(q)}{e^{-\betaeff \hbar\omega_0}+1}G_i\left(\omega_0+q\Omega\right)\left(e^{-\betaeff \hbar\omega_0}-e^{-\beta_i \hbar \left(\omega_0+q\Omega\right)}\right),
\end{equation}
where the notation is the same as in Eqs.~\eqref{eq_J} and~\eqref{eq_betaeff} in the main text.

\end{widetext}

\end{document}